# Analytical model of three regimes of cold cathode breakdown in helium

Liang Xu[1, 2*], Alexander V. Khrabrov[2], Igor D. Kaganovich[2], Timothy J. Sommerer[3]

[1]*Department of Modern Physics, University of Science and Technology of China, Hefei, Anhui 230026, China*

[2]*Princeton Plasma Physics Laboratory, Princeton, NJ 08543, USA*

[3]*General Electric Research, Niskayuna, NY 12309, USA*

E-mail: hfxuliang@gmail.com; ikaganov@pppl.gov

**Abstract.** An analytical model has been developed to map out the low-pressure (left-hand) branch of the Paschen curve at very high voltage when electrons are in the runaway regime and charge exchange/ionization avalanche by ions and fast neutral atoms becomes important. The model has been applied to helium gas between parallel-plate electrodes, at potentials ranging in magnitude between 10 and 1000 kilovolt. The respective value of reduced electric field $E/n$ varies in the range of $50-6000 \, \mathrm{kTd}$ ( $1 \, \mathrm{kTd} = 10^{-18} \, \mathrm{Vm}^{-2}$ ), with reduced density $nd$ (where $n$ is the gas density and $d$ is the inter-electrode distance) on the order of $10^{20} \, \mathrm{m}^{-2}$. For fast neutral atoms produced via charge exchange, the following interactions prove essential to understanding the breakdown mechanism: contribution to impact ionization, strongly anisotropic scattering in collisions with background atoms, and backscattering (of both atoms and ions, which become neutralized) from the cathode. Three regimes of the breakdown have been identified according to relative share of impact ionization by electrons, by ions, and by fast neutrals. In the fast-neutral regime of particular interest here, the ionization avalanche is directed from anode towards cathode. It is initiated by the fast neutral beam backscattered from the cathode, and charge multiplication occurs through multiple successive cycles of ionization and charge exchange. Further, the double-valued shape of the Paschen curve with a branch point near 200 kV is found to be due to heavy species undergoing a transition to runaway regime. The analytical Paschen curve is compared to those obtained with a detailed Particle-In-Cell/Monte Carlo (PIC/MCC) simulation, and also to experimental measurements [11]. The model provides accurate predictions for $E/n$ up to $\sim 10^3 \, \mathrm{kTd}$, constrained by availability and quality of required input data.

---

[*] Current address: *Department of Electrical Engineering and Information Science, Ruhr University Bochum, D-44780 Bochum, Germany*

# 1. INTRODUCTION

Understanding the mechanism of ionization breakdown in gases has been a major effort in discharge physics for more than a century. For parallel-plate DC discharge, the breakdown condition is known as the Paschen law [1], according to which the breakdown voltage *V* is a function of the product of the gas pressure *p* (or density *n*) and the electrode separation *d*. When the applied voltage is low (say, below 1 kV), the specific relation between *V* and *pd* assumes a simple analytical form. It results from applying a local-field theory to electron-impact ionization and assuming a constant yield for ion-induced electron emission [2]. However, under the conditions of high voltage/low pressure on the left-hand branch of the Paschen curve, additional elementary processes associated with fast ions and fast neutrals produced in charge transfer become necessary for explaining the experimental data [3, 4, 5, 6, 7, 8, 9, 10].

Previously in [11], the authors presented a Paschen curve for helium, predicted by PIC/MCC simulations of Townsend discharge. The studied range of applied voltage was 100-1000 kV. It was found, in particular, that anisotropic scattering of all particle species on the background neutrals and backscattering of fast neutrals (including ions, neutralized upon impact) at the cathode surface were essential for identifying the breakdown state, and had to be accounted for. The Paschen curve predicted by the kinetic model is double-valued, with *V(pd)* having a turning point at approximately 300 kV. This feature indicates an essential role of heavy species (ions and fast neutrals). It is due to the onset of the runaway condition, analogously to the violation of locality for electrons at much lower voltage (kilovolt range in the case of helium). Direct particle simulation, with adequate models and input data for gas-phase and surface interactions, allows predicting the properties of high-voltage discharge with minimal assumptions. At the same time, it would be of advantage to have a reduced model for which much of the analysis could be carried out analytically. Such model would provide basic insight into the nature of the process, facilitating both experimental work and verification of the kinetic numerical model.

In theoretical models of gas breakdown, most authors [12, 13, 14] only consider the electron-impact ionization source, and the emission of secondary electrons from the cathode due to bombardment by ions, photons, and metastable atoms. This is generally sufficient for $E/n$ below 1 kTd. Efforts to account for the roles of heavy particles, particularly in argon and helium, have been underway for several decades. Jelenković *et al*. [15] investigated the Paschen curve in helium for $E/n$ in the range 0.3 - 9 kTd within an electron/ion/fast-neutral model. Local-field equilibrium was assumed for all three particle species, to yield results consistent with experimental data for $E/n$ in the range in question. Phelps *et al*. [16] established a model for breakdown in argon that accounted for most of the physical processes, also based on local-field equilibrium distributions of electrons, ions and fast neutrals. The Paschen curve predicted by those authors was in excellent agreement with experimental data for *V* < 3 kV, or approximately for $E/n$ < 100 kTd. On the other hand, as demonstrated by Monte

Carlo simulation [17], in helium electrons undergo a transition to runaway regime when *E/n* is greater than 850 Td. To address the non-local behavior of electrons in the case of argon, Phelps *et al.* [18] applied a "single beam", or mono-energetic electron beam model and obtained a Paschen curve consistent with experimental results up to 1000 kTd, with a fixed value of ion-induced secondary electron yield $\gamma_i = 0.05$. Early on, Granzow *et al.* [19] developed a theoretical model for the low-pressure branch of the Paschen curve for D$_2$ gas for voltages ranging between 5 and 120 kV. They accounted for charge exchange and ionization by ions and fast neutrals, as well as for backscattering of electrons from the anode which was found to be important. Macheret and Shneider [10], also investigating breakdown in argon, utilized a "forward-back" (two-beam) approximation for electron velocity distribution. Those authors also identified the important role of ionization due to accelerated ions and to fast neutrals produced in charge exchange. However, backscattering of ions (which mostly neutralize) and fast neutrals at the cathode surface is essential at *V*>100 kV and needs to be accounted for in any applicable theory. Also, not much analysis has been done in which energy dependence of charge exchange cross-section as well as anisotropy in heavy-particle collisions were both accounted for.

Based on the results of the particle simulations presented in our previous paper [11], the present work aims to develop a realistic analytical model of gas breakdown in helium for *E/n* on the order of 1000 kTd. Such model, incorporating electron, ion and fast neutral species, has been formulated to properly describe their interactions with the background gas and with the electrode surfaces. Particular attention has been paid to the significant role of anisotropic scattering experienced by fast atoms and to the particle backscattering at the electrodes, as well as to ionization by fast-neutral impact. The observed double-valued behavior of the Paschen curve, with the turning point at about 200 kV, will be revisited. The specific details of the analytical model are given in Section **2** and the results are presented and discussed in Section **3**. Section **4** **s**ummarizes the work.

## 2. Analytical model

The model accounts for kinetics of positive ions, fast neutral atoms, and electrons in a Townsend discharge in helium at extremely high values of $E/n$. The underlying mechanisms are charge transfer that controls the velocity distribution of the ions, and free-fall motion of electrons whose free path exceeds the electrode separation [11]. The high voltage breakdown criterion is defined as the marginal condition where self-sustained steady state is found, with a balance between total (per unit area) ionization rate and net fluxes of ions (or electrons) through the boundaries.

For the reduced electric field *E/n* on the order of 1000 kTd, velocity distributions of ions transported towards the cathode and of electrons transported towards the anode are strongly peaked near the direction of the electric field. Same applies to fast neutral atoms resulting from charge exchange, due to strongly anisotropic scattering on the background neutrals. The respective velocity distributions will be treated as one-dimensional. Cosine distributions will be assumed for electrons backscattered at the anode and for fast neutrals backscattered at the cathode (meaning half-isotropic velocity distributions of the backscattered species in the discharge volume). In the analytical model, the elementary processes responsible for gas-phase ionization are electron-, ion-, and fast-atom impact. The surface interactions include ion- and fast-neutral-induced secondary electron emission (SEE) from the cathode, and fast-neutral backscattering from the cathode. Ion backscattering from the cathode (as neutrals) and fast-neutral backscattering from the anode, which were both accounted for in the particle model [11], will be neglected due to their fluxes being very small compared to the primary (produced by charge transfer) fast-neutral flux collected at the cathode. Photon-induced secondary electron emission, which was verified to be a negligible process in the PIC/MCC model, is also disregarded. Inelastic electron backscattering at the anode plays an important role in gas breakdown [5, 20, 21] and will be accounted for in the effective "beam" description of the process. We note that all sets of data for energy-dependent cross sections, backscattering coefficients, and secondary electron yields are identical with those in [11], in order to compare the results with those obtained by particle simulations. In our model, we use notation $\Gamma_x$ to represent absolute value of the respective flux and the direction from cathode to anode is chosen as positive.

## 2.1. Electron model

For *E/n* in the range of interest for the present study, electrons are in the runaway regime, as already known and verified in particle simulations presented in our preceding work [11]. Fig. 1 shows examples of electron velocity distributions (plotted vs. energy; without backscattered contribution) at the anode. Electrons emitted from the cathode impact the anode as ballistic beams. Electrons produced by ionization in the volume are also free-falling, and the shape of their distribution (the tail) is determined by the profile of the ionization rate.

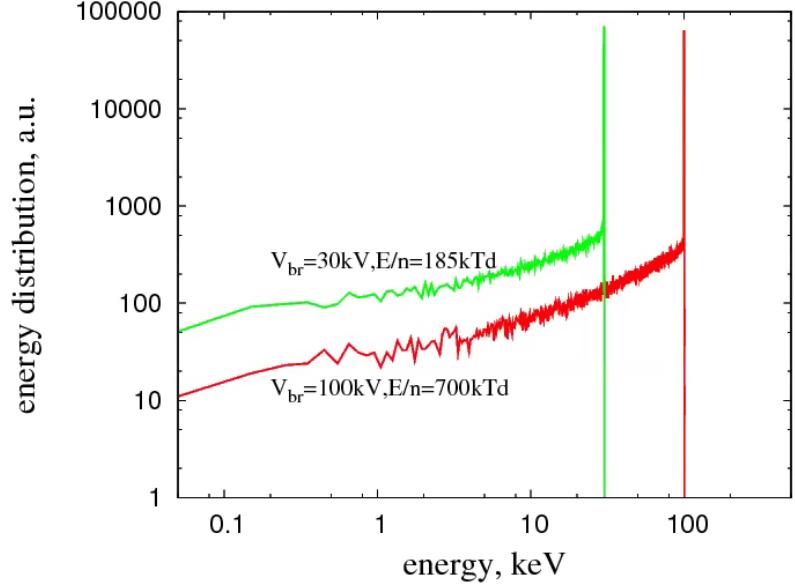

Fig. 1: Electron velocity distributions (flux-energy distributions) at the anode found in Monte Carlo simulations at two discharge voltages of 30 kV and 100 kV. It is seen that in both cases electron beams are formed in the gap. The low-energy tails are due to electrons produced by impact ionization of the gas by electrons, ions and fast neutrals. The electrons undergo a free fall and their energy spectrum corresponds to the profile of the ionization rate.

Under the free-fall approximation, the spatial ionization coefficient of electrons $\alpha_{ei}(x)/n$ can be approximated based on a single-beam model [22]:

$$\frac{\alpha_{ei}(x)}{n} = \frac{\int_0^\infty Q_{ei}(\varepsilon)\delta(\varepsilon - eEx)d\varepsilon}{\int_0^\infty \delta(\varepsilon - eEx)d\varepsilon} = Q_{ei}(\varepsilon = eEx), \qquad (1)$$

where $Q_{ei}(\varepsilon)$ is electron-impact ionization cross section and $x=0$ at the cathode. We follow the terminology of [15, 18] where $\alpha/n$ is called "spatial" reaction (e. g., ionization, charge-exchange, etc.) coefficient. We note that in the extremely high electric field, electrons backscattered from the anode have important effect on electron ionization coefficient [5, 20, 21, 23, 24]. These electrons become trapped (possibly undergoing repeat reflections) and contribute to ionization as they slow down bouncing off the potential barrier. The particle model [11] allowed for a realistic angular distribution, namely the cosine law, of the backscattered flux, with the backscattering probability $\gamma = 0.028 \times \exp[1.154 \times (1 - \cos\theta)]$ based on experimental data [25]. In the reduced model, we ignore ionization due to electrons released in the gas. Therefore the ionization by electrons is attributed to the impact of primary (cathode-emitted) electrons and to the electrons backscattered (possibly more than once) at the anode. Becasue the backscattered electrons are trapped, they do not contribute to the net flux which is still given by that of the primary electrons. At the same time, the ionization rate due to backscattered electrons will be still proportional to the primary flux and so will be the total electron-impact ionization rate.

In Fig. 2(a), we show a profile of the spatial ionization coefficient by electrons as deduced from our Monte Carlo simulations, and compare it to that predicted by simulations with no account for electron backscattering. As expected, the difference between the former and the latter increases towards the anode. At the anode surface, it is about a factor of 3. The backscattered electrons which belong to the low-energy portion of the spectrum will be trapped in the vicinity of the anode (especially considering that each successive reflection involves an energy loss). Therefore the profile of the volume ionization coefficient due to backscattered electrons will be inverted relative to that due to the accelerating cathode-emitted primaries. Actually, Fig. 2(a) indicates that the ionization coefficient in the particle model can be treated as spatially independent over the discharge gap, save to the vicinity of the cathode, where low-energy primary electrons dominate (given that the maximum of electron-impact ionization cross-section occurs at energy several times the ionization threshold; about 130 eV for helium). The weakly varying spatial ionization coefficient is also observed at other values of the breakdown voltage, as seen in Fig. 2(b) where three cases are presented.

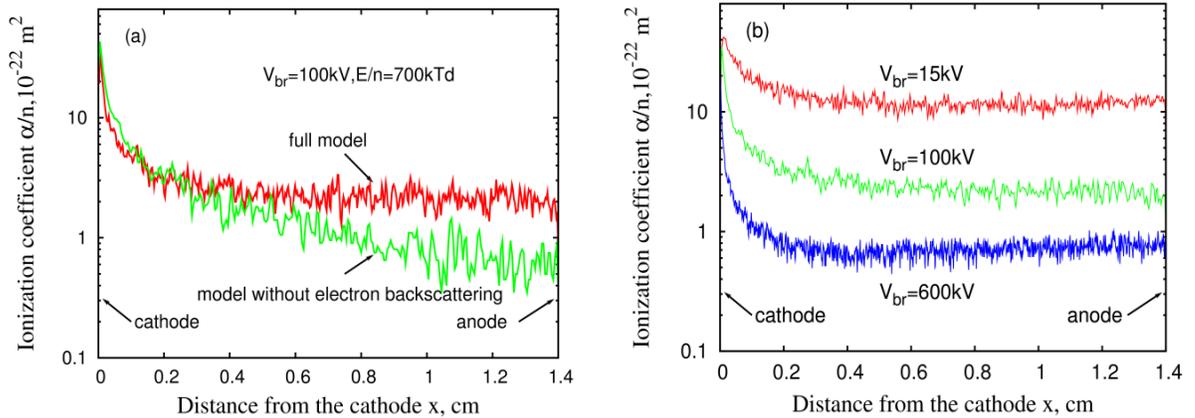

Fig. 2: Ionization coefficients $\alpha/n$ for electrons found in Monte Carlo simulations [11]: (a) spatial ionization coefficients obtained in the full model and the model without electron backscattering, for breakdown voltage of 100 kV, and (b) spatial ionization coefficients obtained for breakdown voltages of 15 kV, 100 kV, and 600 kV.

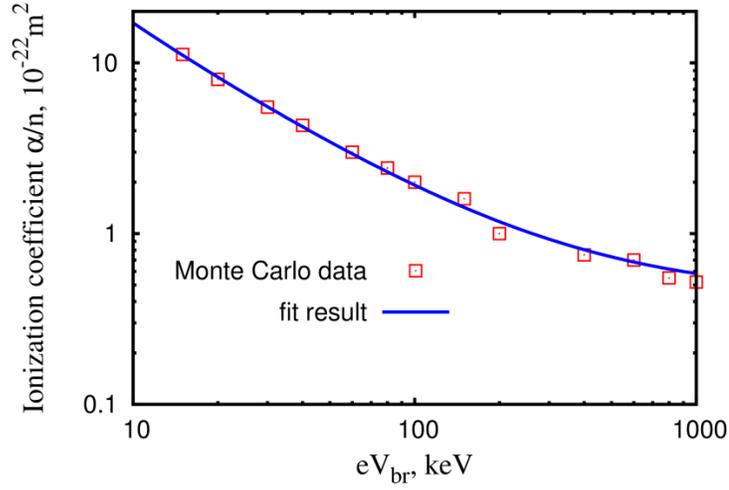

Fig. 3: Analytical fit for ionization coefficient versus $eV_{br}$ with account for both primary and backscattered electron fluxes.

If, as discussed above, the electron-ionization reaction coefficient can be reasonably approximated by a constant value in the entire discharge gap, the description of non-local electron kinetics becomes simplified and the ionization coefficient can be treated as effectively "local". In Fig. 3 the constant-value approximations of electron-induced ionization coefficient $\alpha_{ei}/n(E,n)$, obtained in PIC/MCC simulations for breakdown voltage 15 kV < $V_{br}$ < 1000 kV, are plotted versus $eV_{br}$. This data allows a simple analytical fit:

$$\frac{\alpha_{ei}(x)}{n} \simeq \frac{\alpha_{ei}}{n}(E,n) = 1.0 \times 10^{-22} \times \left[ 0.436 + 1.468 \times 10^5 / (eV_{br} - 1206.8) \right], \qquad (2)$$

where $V_{br}$ in Volts. The value of $\alpha_{ei}$ can be now used in the fluid equation for electron flux in the steady state:

$$\frac{d\Gamma_e(x)}{dx} = \alpha_{ei}(E,n)\Gamma_e(x) + \alpha_{ii}(E/n)\Gamma_i(x) + \alpha_{ai}(E/n)\Gamma_a(x) + \alpha_{bi}(E/n)\Gamma_b(x), \qquad (3)$$

where $\Gamma_e(x), \Gamma_i(x), \Gamma_a(x), \Gamma_b(x)$, respectively, are the particle fluxes of electrons, ions, primary fast atoms, and backscattered fast atoms. Therefore the electron flux is increasing toward the anode due to the combined ionization processes. The quantities $\alpha_{ii}/n(E/n), \alpha_{ai}/n(E/n), \alpha_{bi}/n(E/n)$ are ion-, fast-neutral-, and backscattered fast-neutral-impact ionization coefficients in helium. The values of coefficients $\alpha_x/n$ for ions and fast atoms which are discussed in Appendix A. Electron-impact ionization of helium is the only process with significant contribution when E/n < 4.3 kTd [9]. Other contributions to the electron particle balance, e. g. electron–ion recombination or associative ionization, can be neglected for the low-current Townsend discharge.

## 2.2. Ion model

The model for He$^+$ ions is based on local equilibrium distribution governed by charge transfer [18, 26], i.e. it is being assumed that the free path $\lambda_{cx}$ for charge transfer is much smaller than the discharge gap. This assumption was verified in the Monte Carlo study of breakdown in helium [11] according to which the charge-transfer free path $\lambda_{cx}$ is one order lower than the discharge gap even in the 1000 kTd range. We note that charge exchange collisions present a friction mechanism in the momentum transfer equation, but do not affect the flux continuity. In addition, the net current in a steady state should be conserved:

$$\Gamma_e(x) + \Gamma_i(x) = const \quad . \tag{4}$$

Therefore due to the ionization of the gas by electrons, ions and fast neutrals, the ion flux increases towards the cathode and obeys the following equation:

$$\frac{d\Gamma_i(x)}{dx} = -\frac{d\Gamma_e(x)}{dx} = -\alpha_{ei}(E,n)\Gamma_e(x) - \alpha_{ii}(E/n)\Gamma_i(x) - \alpha_{ai}(E/n)\Gamma_a(x) - \alpha_{bi}(E/n)\Gamma_b(x) \quad . \tag{5}$$

Accounting only for charge transfer, the dominant collision type for ions, we solve the resulting steady-state Boltzmann equation. The charge exchange cross section approximates as $\sigma_{cx}(\varepsilon) = [A - B\ln\varepsilon]^2 \times \sigma_0$, where $A = 5.282, B = 0.294, \sigma_0 = 10^{-20} m^2$ and energy $\varepsilon$ is in eV. The velocity distribution, as a function of energy, is

$$f(\varepsilon) = C \exp\left[-\varepsilon \frac{A^2 + B^2 \ln^2\varepsilon - 2(AB + B^2)(\ln\varepsilon - 1)}{E/n\sigma_0}\right]. \tag{6}$$

where C is the normalization factor defined by the flux. We note, however, that the Boltzmann solution (local approximation) shows large deviation from the ion spectrum observed in Monte Carlo simulations, as seen in Fig. 4 for $V$=100 kV. The Boltzmann solution predicts a "hotter" tail. This indicates that ionization frequency can be comparable to charge exchange frequency, to results in large amount of slow ions released in the ionizing collisions. This effect will be studied in future work. Another option is to adopt a one-dimensional Maxwellian distribution with "temperature" $T_i$ on the order of $eE\lambda_{cx}$ to approximate the ion velocity distribution as $f_i(\varepsilon) = C' \exp(-\varepsilon/T_i)$, where $C'$ is again the defined by the flux. This approach (Maxwellian approximation) was adopted by Phelps *et al.* [27] for high values of *E/n*. The empirical fit for ion temperature vs. *E/n* is based on experimental data:

$$T_i = 4\left[(E/n)/1000\right]^{1.2}, \tag{7}$$

where $E/n$ is in Td and $T_i$ is in eV. In Fig. 4, the above approximation is also compared to the ion energy distribution function yielded Monte Carlo model for breakdown voltage of 100 kV. It is found that the Maxwellian approximation shows good agreement with the Monte Carlo result, except at very low energies (<3keV). The deviation at low energies should have negligible effect on the ionization coefficient, the dominant contribution to which comes from high-energy ions.

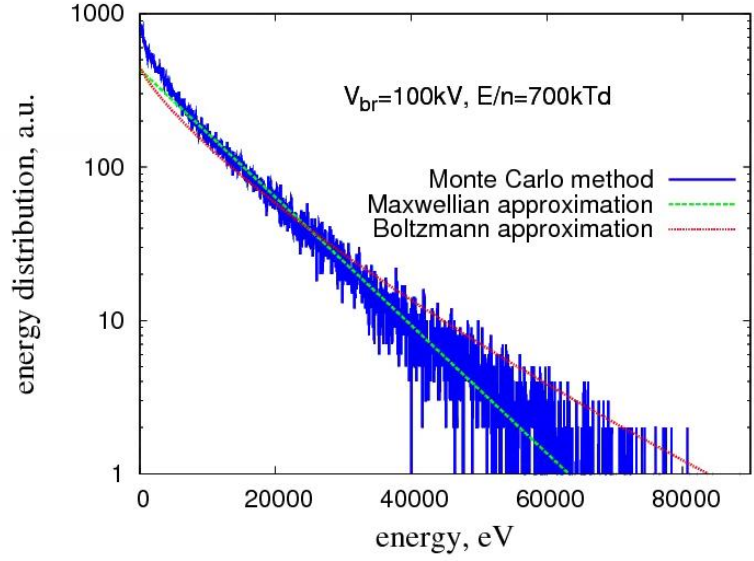

Fig. 4: Ion velocity distribution function, vs. energy, measured at the cathode in Monte Carlo simulations [11] at breakdown of voltage 100 kV, compared with the Maxwellian approximation with the local approximation based on solving the Boltzmann equation. It is interesting to observe that the Maxwellian approximation shows good agreement with the Monte Carlo result over much of the energy range, except at low energies below 3 keV, whereas the local Boltzmann solution predicts a hotter tail. This phenomenon is attributed to "loading" of the distribution with newly produced slow ions. The detailed interpretation will be addressed in our future work.

Next, the charge transfer coefficient $\alpha_{ct}/n$ and the ionization coefficient $\alpha_{ii}/n$ for $He^+$ in neutral helium are plotted versus the reduced electric field $E/n$ in Fig. 5. These coefficients are obtained by flux-averaging the respective cross-sections over the one-dimensional ion distribution and the details are given in Appendix A. The same cross sections adopted in Ref. [11] are utilized in the present work.

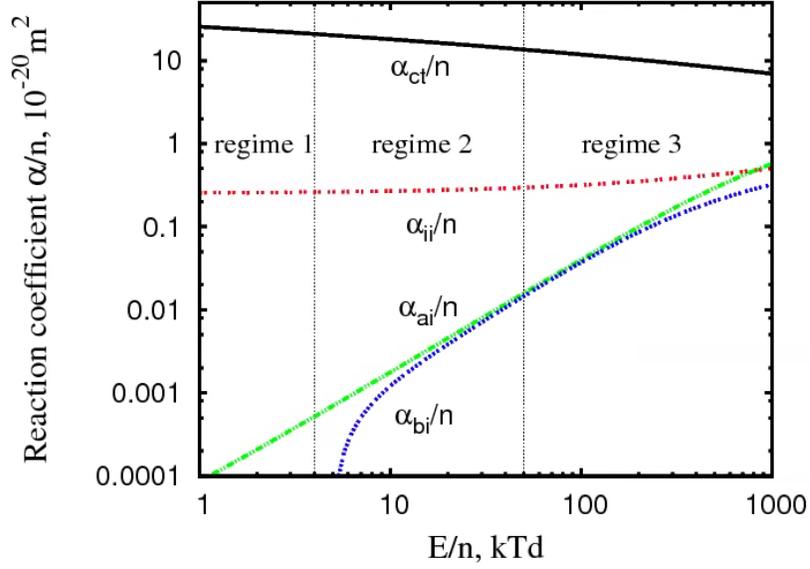

Fig. 5. Reaction coefficients for He$^+$ and fast helium atoms (primary and backscattered) in background helium gas. Only those reactions that are essential to creating an ion-fast-neutral-ion avalanche are taken into account. For ions, only charge transfer ($\alpha_{ct}/n$) and ionization ($\alpha_{ii}/n$) processes are considered. For primary and backscattered fast neutrals, only the ionization reactions ($\alpha_{ai}/n$ or $\alpha_{bi}/n$) are taken into account. The values of the coefficients are discussed in Appendix A. The regimes are defined in Table 1.

Regarding the surface interactions, the analytical model accounts for secondary electron emission (SEE) induced by both ions and fast neutrals. Therefore the electron flux leaving the cathode can be expressed as following:

$$\Gamma_e(0) = \gamma_i(E/n)\Gamma_i(0) + \gamma_f(E/n)\Gamma_a(0) \ . \qquad (8)$$

where $\gamma_i, \gamma_a$ are the *E/n*-dependent secondary yields induced, respectively, by ions and by fast neutrals. The yield coefficients that relate net fluxes are obtained by averaging the respective energy-dependent SEE yields over the ion and corresponding primary fast-atom distributions. The resulting yield coefficients (surface reaction rates) are discussed in appendix A and plotted in Fig. 6.

The boundary condition for ions at the anode states that there are no accelerated ions; the same applies to fast neutrals:

$$\Gamma_i(d) = \Gamma_a(d) = 0. \qquad (9)$$

Ions emitted from the anode due to electron bombardment are ignored, as is the tertiary flux of backscattered fast neutrals.

## 2.3. Fast neutral atom model

Fast atoms are generated primarily via charge transfer and therefore their velocity distribution is governed by that of the projectile He$^+$ ions [18, 26]. In studies of high-voltage discharges, e.g. [10, 15, 16], the usual assumption has been that any elastic or inelastic collision of a fast neutral with a background atom of equal mass would result in a loss of that fast neutral, by reducing its energy below the threshold of excitation or ionization. However, the situation is different with anisotropic scattering which occurs at ion/atom energies corresponding to high values of *E/n*. Energy loss by fast atoms can be negligible in elastic and excitation collisions due to small scattering angles. The main loss channel for fast neutrals will be through ionization, because stripping of the projectile and ionization of the target occur with equal probability. Therefore the equation for the fast neutral flux traveling towards the cathode (in the negative direction) can be written as follows:

$$\frac{d\Gamma_a(x)}{dx} = -\alpha_{ct}(E/n)\Gamma_i(x) + \frac{1}{2}\alpha_{ai}(E/n)\Gamma_a(x). \qquad (10)$$

The ionization coefficient $\alpha_{ai}/n$ in Eq. (10) is obtained by averaging the neutral-impact ionization cross section $Q_{ai}(\varepsilon)$ over the fast-neutral energy distribution. For this purpose, assuming the first term in (10) dominates over the second, first approximation for the neutral distribution will suffice, namely $f_a(\varepsilon) \propto Q_{ct}(\varepsilon)f_i(\varepsilon)$. The result is

$$\frac{\alpha_{ai}}{n}\left(\frac{E}{n}\right) = \frac{\int_0^\infty Q_{ai}(\varepsilon)f_a(\varepsilon)d\varepsilon}{\int_0^\infty f_a(\varepsilon)d\varepsilon} = \frac{\int_0^\infty Q_{ct}(\varepsilon)Q_{ai}(\varepsilon)f_i(\varepsilon)d\varepsilon}{\int_0^\infty Q_{ct}(\varepsilon)f_i(\varepsilon)d\varepsilon}. \qquad (11)$$

The slowly decreasing $Q_{ct}(\varepsilon)$ results in a "cooler" distribution of fast atoms versus that of the ions. At the cathode, primary fast neutrals will be backscattered and travel freely against the electric field, which is essential for initiating the heavy-species ionization avalanche in the vicinity of the anode. As is the case for primary fast atoms, stripping is the prevailing elimination mechanism for the backscattered neutrals. Therefore for the backscattered fast-atom flux, dubbed $\Gamma_b$, we have

$$\frac{d\Gamma_b(x)}{dx} = -\frac{1}{2}\alpha_{bi}(E/n)\Gamma_b(x), \qquad (12)$$

where $\alpha_{bi}/n$ is the flux-averaged ionization coefficient for backscattered fast neutrals, calculated by averaging the fast atom ionization cross section $Q_{ai}(\varepsilon)$ over the backscattered fast-atom energy distribution $f_b(\varepsilon) \propto R_N(\varepsilon)f_a(\varepsilon) \propto Q_{ct}(\varepsilon)R_N(\varepsilon)f_i(\varepsilon)$. Becasue a good model for the energy spectrum of backscattered flux is not available, the energy of a reflected atom is calculated as $\varepsilon' = \varepsilon\dfrac{R_E(\varepsilon)}{R_N(\varepsilon)}$, where $R_N(\varepsilon)$ and $R_E(\varepsilon)$ are, respectively, the particle flux and the energy flux reflection coefficients for the primary atom energy $\varepsilon$. Also, a cosine angular distribution is adopted for the backscattered flux, same as in [11]. The resulting expression for backscattered-fast-neutral ionization coefficient is

$$\frac{\alpha_{bi}}{n}\left(\frac{E}{n}\right) = \frac{\int_0^\infty R_N(\varepsilon) Q_{ct}(\varepsilon) Q_{ai}(\varepsilon') f_i(\varepsilon) d\varepsilon}{\int_0^\infty Q_{ct}(\varepsilon) R_N(\varepsilon) f_i(\varepsilon) d\varepsilon} \times \frac{\int_0^{\pi/2} \cos\theta \sin\theta d\theta}{\int_0^{\pi/2} \cos^2\theta \sin\theta d\theta}. \quad (13)$$

where the first factor on the right is similar to Eq. (11) and the second factor is 3/2. The calculated $\alpha_{bi}/n$ is plotted in Fig. 5. It is seen that even with account for angular distribution, values of $\alpha_{bi}/n$ are always smaller than those of $\alpha_{ai}/n$, indicating the importance of energy loss in inelastic backscattering.

Next, to formulate the cathode boundary condition for the fast neutral flux, we need an expression for the reflection coefficient $R_a(E/n)$. It is obtained by averaging the energy-dependent particle backscattering coefficient $R_N(\varepsilon)$ over the primary fast-atom distribution $f_a(\varepsilon)$:

$$R_a\left(\frac{E}{n}\right) = \frac{\int_0^\infty Q_{ct}(\varepsilon) R_N(\varepsilon) f_i(\varepsilon) d\varepsilon}{\int_0^\infty Q_{ct}(\varepsilon) f_i(\varepsilon) d\varepsilon}. \quad (14)$$

This coefficient is plotted in Fig. 6 and its calculation is given in Appendix A. Therefore the backscattered fast neutral flux at the cathode is given by

$$\Gamma_b(0) = R_a \Gamma_a(0). \quad (15)$$

Here, we note that in the present model, the ions neutralized and then backscattered as neutrals at the cathode are neglected, due to their flux being much smaller than the primary fast-neutral flux. The fast atoms repeatedly backscattered at the anode are likewise neglected, because their flux is on the order of $R_a^2 \ll 1$ and their energy spectrum also degrades upon successive reflections.

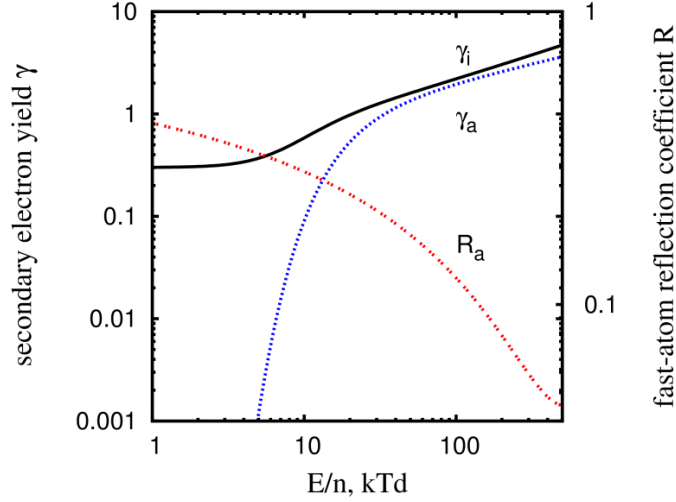

Fig. 6. Surface interaction coefficients: secondary electron yield due to ions ($\gamma_i$) and to fast atoms ($\gamma_a$), and fast-atom backscattering coefficient ($R_a$). The values of these coefficients are considered in Appendix A.

## 3. Results and discussion

### 3.1. The physics of ionization avalanche

To gain insight into how the avalanche is sustained in very high electric field, we consider spatial profiles of particle fluxes predicted by the analytical model for the breakdown state. Fig. 7 shows the calculated fluxes of electrons, ions and fast neutrals, compared with those obtained in a Monte Carlo simulation of a breakdown state for $V_{br} = 100 kV$, with corresponding gas density $n = 1.018 \times 10^{22} m^{-3}$. The particle fluxes are normalized to the constant total flux $\Gamma_t = \Gamma_i + \Gamma_e$ and the calculation details are given in Appendix B. As seen in Fig. 7, the calculated electron and ion fluxes are in excellent agreement with the particle simulation results. The relative change in the electron flux over the gap is small because the secondary yield at the cathode due to energetic neutral flux (about 10) is much higher than the multiplication factor (which is correspondingly about 1.1). We note that in both analytical and numerical Monte Carlo models, the net flux of fast neutrals $(\Gamma_a - \Gamma_b)$ is negative within some distance of the anode, because $\Gamma_a(d) = 0$ in the former and $\Gamma_a(d) = R\,\Gamma_b(d)$ in the latter, with $R \ll 1$. The deviation between net fast-neutral fluxes $(\Gamma_{a,cal} - \Gamma_{b,cal})$ of the reduced model and $(\Gamma_{a,mcc} - \Gamma_{b,mcc})$ of the particle model is, for the most part, due to low-energy cutoff applied to fast neutral species in the Monte Carlo code. The cut-off is at energy of 200 eV where ionization frequency becomes negligible.

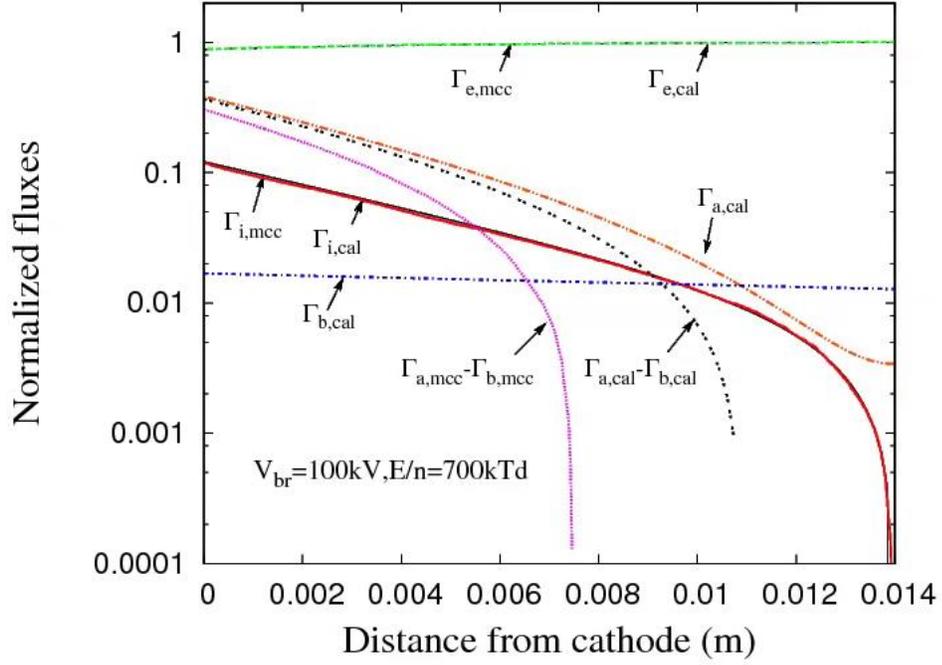

Fig. 7 Spatial profiles of normalized fluxes of electrons $\Gamma_{e,cal}$, He$^+$ ions $\Gamma_{i,cal}$, primary fast atoms $\Gamma_{a,cal}$, and reflected fast atoms $\Gamma_{b,cal}$, compared to the corresponding fluxes $\Gamma_{e,mcc}, \Gamma_{i,mcc}$ and $\left(\Gamma_{a,mcc}-\Gamma_{b,mcc}\right)$ obtained with Monte Carlo model [11] at breakdown voltage of 100 kV. Electron flux and ion flux are seen to be in excellent agreement between the analytical model and Monte Carlo model, while the net fast atom flux $\left(\Gamma_{a,cal}-\Gamma_{b,cal}\right)$ does not agree with Monte Carlo result $\left(\Gamma_{a,mcc}-\Gamma_{b,mcc}\right)$. The discrepancy is due to the fast-neutral species definition in our Monte Carlo model: only those atoms with energies above 200 eV are tracked as "fast neutrals". The particle flux calculation is discussed in Appendix B.

In order to identify the roles of individual particle species in sustaining the charge multiplication avalanche, in Fig. 8 we plot the profiles of gas ionization rates due to electrons, ions, primary fast atoms, and backscattered fast atoms for the case when the breakdown voltage is 100 kV. Note that the ionization rate is, again, normalized by the net charge flux $\left(\Gamma_e+\Gamma_i\right)$. It is clearly seen that primary fast neutrals make the largest contribution to the ionization rate, and it increases from anode towards the cathode as the flux of fast neutrals multiplies. Near the cathode, electron-impact ionization rate is much smaller than that due to fast atoms and to ions, but the opposite is true in the vicinity of the anode. On the whole, in the total ionization rate integrated over the gap, the share of electrons among the three species is only 22% and that of fast neutrals (primary + backscattered) is as high as 57%.

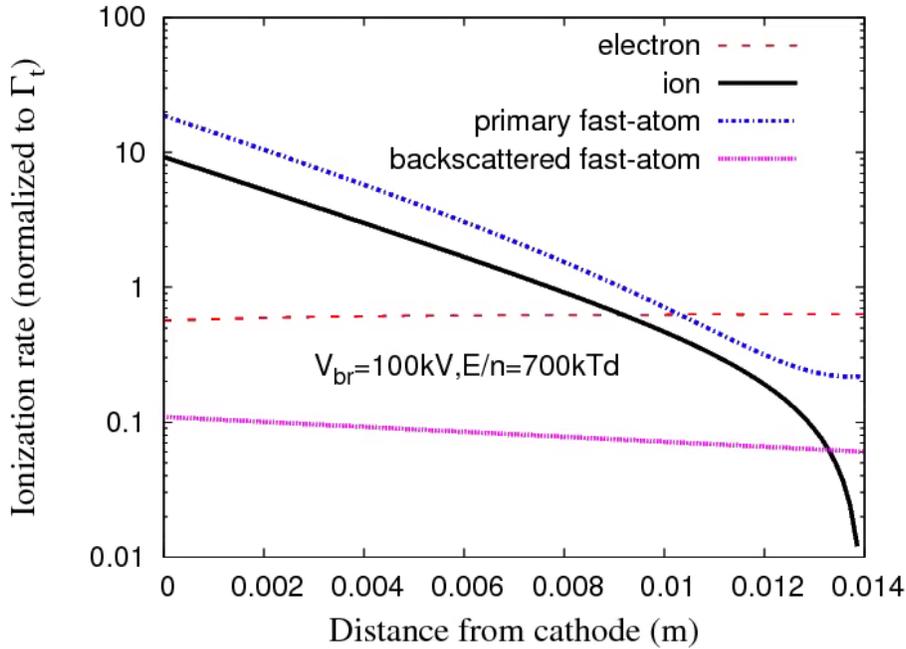

Fig. 8: Ionization rates due to impact by electrons, ions, primary fast atoms, and backscattered fast atoms calculated in the reduced analytical model. Their profiles are determined by the respective fluxes. Overall, the fast-atom contribution to net ionization is the largest and electrons contribute about the same as ions.

### 3.2. Three regimes of breakdown discharge

The structure of the breakdown discharge changes in accordance with the value of the reduced electric field $E/n$ which controls the volume-reaction and surface-reaction (electron emission and backscattering) rates of individual particle species. Therefore it is of interest to identify distinct regimes of the breakdown and discuss transitions from one to another. Presently, we introduce three such regimes, called (1) "electron regime", (2) "ion regime", and (3) "fast-atom regime". The ordering corresponds to increasing $E/n$. Based on our model and previous work on electrical breakdown in helium, these regimes can be distinguished qualitatively according to the importance of gas ionization by each of the respective particle species. In regime 1, identified for helium as $E/n < 4$ kTd, the discharge is sustained only by electrons, because fast-atom energies and ion energies are both below than their ionization thresholds. In regime 2, identified as $4$ kTd $< E/n < 50$ kTd, ion-species contribution to the gas phase ionization prevails over those of electrons and fast neutrals. In regime 3, identified as $E/n > 50$ kTd, ionization by all three particle species is important. In Table 1 we identify the roles of various elementary physical processes under the three regimes listed above. In fact, the corresponding portions of Paschen curve can be adequately described by models (presented below) obtained by further simplifying the reduced analytical model under study, although for regimes 1 and 2, local-field model for electrons [2, 15] should be applied instead of the free-flight model considered presently.

Table 1: Gas-phase elementary processes and particle-surface interactions accounted for in reduced models under different regimes of breakdown.

| reaction / regime | Electron ionization | Ion ionization | Fast-atom ionization | Ion induced electron emission | Fast-atom induced electron emission | Fast-atom backscattering |
|---|---|---|---|---|---|---|
| Regime 1 (model 1) | ✔ | | | ✔ | | |
| Regime 2 (model 2) | ✔ | ✔ | | ✔ | (depends on cathode material) | |
| Regime 3 (model 3) | ✔ | ✔ | ✔ | ✔ | ✔ | |
| Regime 3 (model 4) | ✔ | ✔ | ✔ | ✔ | ✔ | ✔ |

In order to highlight the differences between three different regimes 1, 2, and 3, we introduce three reduced models, identified with the aid of Table 1. In each of these models, we disable those elementary processes of the base model which can be neglected. The complete (base) model is labeled "model 4". We note that in model 3, the process of fast-neutral backscattering is also neglected, to observe (and compare with model 4) the effect of such assumption, commonly used in the literature. Therefore the reduced models can be characterized as follows:

1. First reduced model: $\alpha_{ii} = \alpha_{ai} = \gamma_a = R_a = 0$.

2. Second reduced model: $\alpha_{ai} = \gamma_a = R_a = 0$.

3. Third reduced model: $R_a = 0$.

4. Non-truncated model.

Fig. 9 shows normalized particle fluxes of models 1, 2, 3 and 4 for gas pressure at which the breakdown voltage is 100 kV. The calculations of electron and ion fluxes are discussed in detail in Appendix B. Note that in all cases, the electron ionization coefficient $\alpha_{ei}/n$ of Section 2A is still utilized instead of the value obtained through local-field model. This introduces a lower limit on the discharge voltage to which the description applies. From Fig. 9, it is seen that the solutions obtained with models 1, 2, and 3 differ from that of model 4, and become successively closer to it. This observation illustrates the importance of ionization by ion and fast-neutral impact. In addition, for model 3, the absence of fast-atom reflection from the cathode results not only in reduced ionization compared to model 4, but also the disappearance of the applicable self-organization mechanism of the discharge [11]. The backscattered fast atoms ionize the gas in the gap, and the resulting ions undergo multiple cycles of acceleration and charge exchange to re-generate the primary fast atoms beam which impinges upon the cathode. This self-organization mechanism is essential for sustaining the discharge current and initiating the breakdown, even without electron ionization impact in the 100-1000 kV range [11].

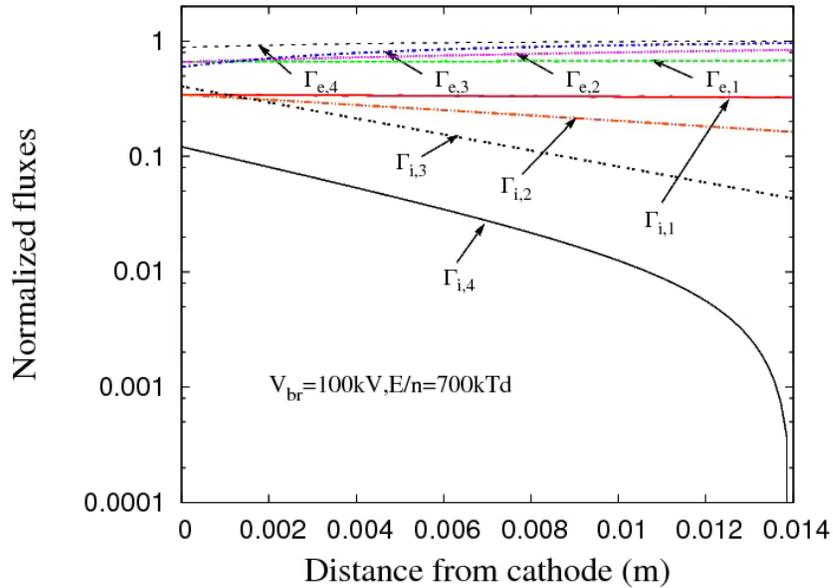

Fig. 9: Calculated profiles of normalized fluxes of electrons, $\Gamma_{e,1}, \Gamma_{e,2}, \Gamma_{e,3}, \Gamma_{e,4}$, and ions, $\Gamma_{i,1}, \Gamma_{i,2}, \Gamma_{i,3}, \Gamma_{i,4}$ for breakdown models 1, 2, 3, 4 specified in the text, for the case in which the breakdown potential is 100 kV. The solutions of the truncated models 1, 2, 3 successively become closer to that of the complete model 4. The difference between models 3 and model 4 indicates that fast atom backscattering is crucial in high-voltage breakdown. The particle flux calculation is discussed in detail in Appendix B.

For the regime 3 (the fast-neutral regime) of interest in this work, in Fig. 10 we visualize the relative shares of impact ionization processes by different species (electrons, ions, and fast neutrals) in the net ionization rate. The results produced by the analytical model are compared with those of with PIC/MCC simulations. Good agreement is obtained between the two sets of calculations. It is seen that in regime 3, electron impact ionization is no longer the dominant process in the production of ions and decreases monotonically with increasing $E/n$. The share of ion-impact ionization increases 18% to 30% over the investigated range of $E/n$. The share of fast-atom impact ionization increases more sharply and becomes the largest when $E/n$ exceeds 300 kTd.

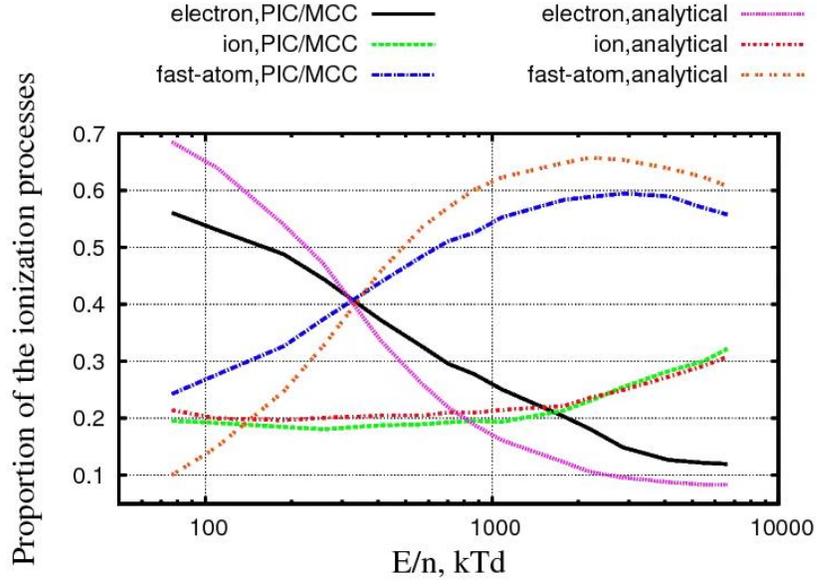

Fig. 10: Fractions of gap-integrated ionization rate due to electron-impact ionization, ion-impact ionization, and fast-atom-impact ionization.

### 3.3. Analytical solution for breakdown threshold

The breakdown threshold is defined as a condition for existence of a non-trivial solution for the set of time-independent particle balance equations (3), (5), (10), and (12) (refer to model 4 in Table 1). Such condition corresponds to a functional dependence between $E/n = V/nd$ and $nd$, i.e. the Paschen curve. The linear equations (3), (5), (10) and (12) are solved subject to appropriate boundary conditions, given by equations (8), (9), and (15). The resulting compatibility condition (an implicit equation for the Paschen curve), necessary for a non-trivial solution, can be expressed as follows:

$$A\exp(\lambda_1 d) + B\exp(\lambda_2 d) + C\exp(\lambda_3 d) + D\exp[(\lambda_1+\lambda_2)d] + E\exp[(\lambda_1+\lambda_3)d] + F\exp[(\lambda_2+\lambda_3)d] = 0 \quad , \tag{16}$$

where $A = \left[R_a(1+\gamma_i)(N-Q) - \beta(1+\gamma_i+N\gamma_a)\right](S-PM)$, $B = \left[R_a(1+\gamma_i)(M+Q) + \beta(1+\gamma_i+M\gamma_a)\right](S-PN)$,

$C = (M-N)(1+\gamma_i)(S-PQ)R_a$,

$D = \left[(QR_a+\beta)(1+P\gamma_i+S\gamma_a) - S(1+\gamma_i+Q\gamma_a)\right](N-M)$,

$E = \left[-NR_a(1+P\gamma_i+S\gamma_a) + S(1+\gamma_i+N\gamma_a)\right](Q-M)$,

$F = \left[MR_a(1+P\gamma_i+S\gamma_a) - S(1+\gamma_i+M\gamma_a)\right](Q-N)$,

$$M = \frac{\alpha_{ct}}{\alpha_{ai}/2-\lambda_1}, N = \frac{\alpha_{ct}}{\alpha_{ai}/2-\lambda_2}, Q = \frac{\alpha_{ct}}{\alpha_{ai}/2-\lambda_3}, P = \frac{\alpha_{ei}}{\alpha_{ii}+2\alpha_{ct}}, S = \frac{2\alpha_{ei}\alpha_{ct}}{\alpha_{ai}(\alpha_{ii}+2\alpha_{ct})} \text{ and}$$

$$\beta = -\frac{\alpha_{ei}}{\alpha_{bi}} + \frac{\alpha_{ii}}{\alpha_{bi}} + \frac{\lambda_3}{\alpha_{bi}} + \frac{\alpha_{ai}\alpha_{ct}}{(\alpha_{ai}/2-\lambda_3)\alpha_{bi}}$$. The four eigenvalues of the matrix corresponding to (3), (5), (10) and (12) are

0 and also $\lambda_{1,2} = \frac{1}{2}\left[(\alpha_{ei}+\alpha_{ai}/2-\alpha_{ii}) \pm \sqrt{(\alpha_{ei}+\alpha_{ai}/2-\alpha_{ii})^2 + 4(\alpha_{ai}\alpha_{ct}+\alpha_{ai}\alpha_{ii}/2-\alpha_{ei}\alpha_{ai}/2)}\right], \lambda_3 = -\alpha_{bi}/2$.

The Paschen curve defined by Eq. (16) is shown in Fig. 11 for 10 kV < V < 1000 kV, alongside the PIC/MCC and experimental results of [11]. The three sets of data are consistent with each other. The analytical model predicts a turning point at about 200 kV, vs. the value 300 kV predicted by the present PIC/MCC model. The respective values of reduced pressure differ by less than 10% when voltage is larger than 50 kV. The discrepancy between the present analytical result and the PIC/MCC model is due to several approximations, e. g., assuming local equilibrium energy distribution of the ions (which is not actually present within several free-path lengths of anode) as well as neglecting ion backscattering at the cathode and fast-neutral backscattering at the anode. Below 50 kV, the large discrepancy observed between the reduced model and the PIC/MCC result indicates that high-voltage electron model fails to adequately describe the electron velocity distribution.

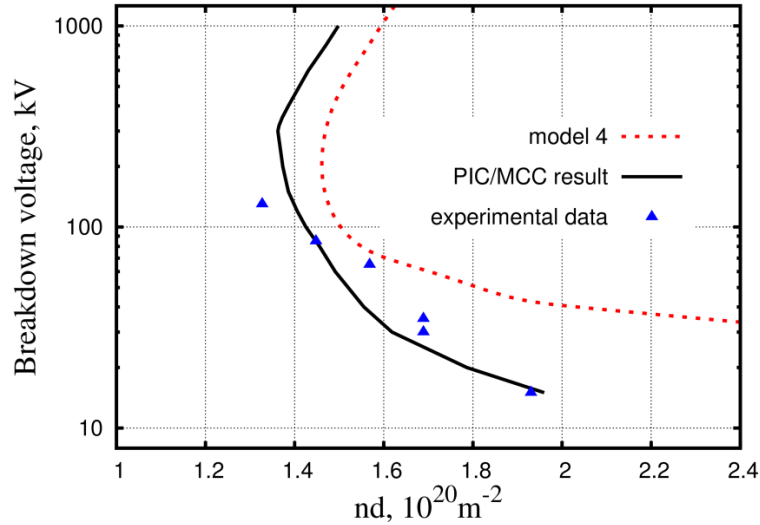

Fig. 11. Paschen curves based on the present analytical calculation being compared with PIC/MCC simulation result, and with experimental data (the latter two from [11]). The experiment covers the range between 15 kV and 130 kV. The analytical model and the PIC/MCC simulation both predict a turning point above 200 kV, i. e., double-valued behavior.

### 3.4. Role of fast-neutral gas ionization and backscattering

To evaluate the significance of fast-neutral backscattering at the electrodes, we obtained the Paschen curve for a model in which this process was disabled. The results are shown in Fig. 12. As expected, absence of backscattered flux causes the Paschen curve to shift to the right. Reducing the backscattering coefficient to zero obviously has a more pronounced effect at lower voltage. This is attributed to the decrease of the backscattering yield with increasing projectile energy. This comparison clearly emphasizes the importance of fast-atom backscattering at the cathode for sustaining a Townsend discharge at high voltage.

We also carried out another calculation, with the fast-atom-impact ionization coefficient reduced by half, for all values of $E/n$, vs. the actual value based on cross-section data. It was done in order to investigate the effect of the fast-atom-impact ionization. The large deviation of the resulting Paschen curve, also shown in Fig. 12, from that based on the un-altered model demonstrates the fast-atom-impact ionization to be an essential at high voltage.

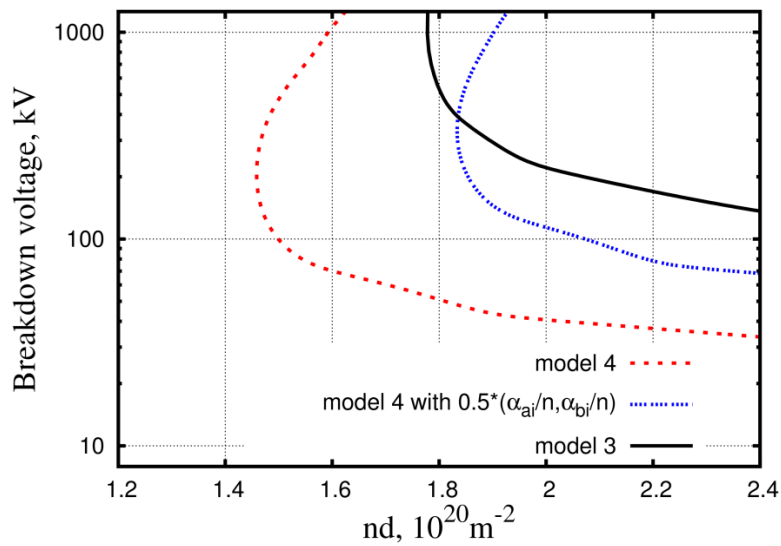

Fig. 12: Comparison of Paschen curves between the full model (model 4), model 4 with fast-atom-impact ionization coefficient reduced by a factor of 2 (for both primary and cathode-backscattered fluxes), and the model with no fast-neutral backscattering at the cathode (model 3). Both modifications to the model cause the Paschen curve to shift strongly to the right, which is indicative of the importance of fast-neutral impact ionization and backscattering.

### 3.5. Effect of fast-atom stripping losses

With strongly anisotropic (i.e. peaked near $0^o$ and $180^o$ in the center-of-mass frame) neutral/neutral scattering built into the model, whenever a fast neutral atom undergoes stripping in an ionizing collision, the target atom remains slow and therefore a fast neutral is lost with a probability of 1/2. To show the importance of this effect, we created another artificial case by eliminating the respective loss term in Eqs. (10) and (12) so that only the background atoms ionize. The resulting

Paschen curve is shown in Fig. 13. It gives much lower values of reduced pressure for given breakdown voltage over the entire range. The extra ionization in the artificial model is due to additional fast neutrals.

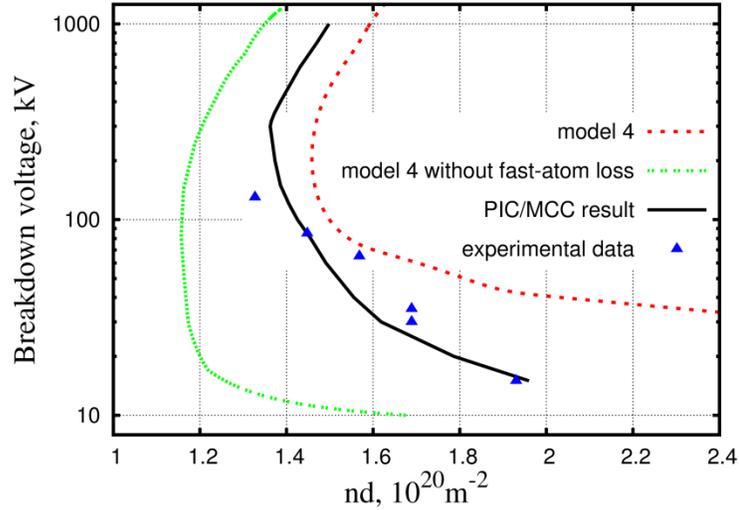

Fig. 13: Paschen curves obtained with the full model (model 4) and in the artificial case without stripping loss of fast-atom, along with the PIC/MCC prediction and experimental results.

### 3.6. Ion-neutral runaway and the turning point on the predicted Paschen curve

The physical mechanism behind the presence of the turning point is that the velocity distributions of the ionizing species, in this case ions and fast neutrals, undergo a transition to the runaway regime. This phenomenon has already been observed for electrons, in which case it occurs at the breakdown voltage in the range of several kiloVolt [9]. Experimentally, such turning points have been observed for helium and mercury [9, 28], and to a lesser extent for neon. Basically, the multiplication length becomes comparable with the electrode spacing and increases monotonically with respect to the applied voltage. Because this length has to fit into the gap, the *nd* value now needs to increase with voltage to have sufficient ionization. We note that the multiplication length in the regime in question can be estimated as $1/\sqrt{\alpha_{ct}\alpha_{ai}}$, based on the expressions for the eigenvalues given above.

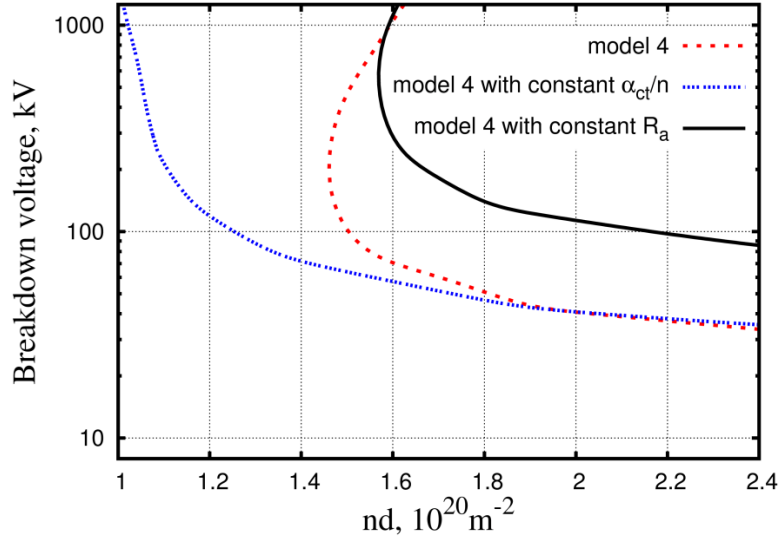

Fig. 14: Paschen curves obtained in two numerical experiments with artificial input data: one with constant charge-transfer cross-section $\alpha_{ct}/n = 1.0\times10^{-19} m^{-2}$, and another with constant backscattering coefficient $R_a = 0.05$ for fast neutrals at the cathode. The un-altered (model 4) result is also shown for comparison.

The decrease in charge transfer coefficient $\alpha_{ct}/n$ with increasing $E/n$ is the primary cause of the runaway transition. To verify this, we performed yet another numerical experiment, with a constant (not depending on $E/n$) charge-transfer reaction coefficient $\alpha_{ct}/n = 1.0\times10^{-19} m^{-2}$. The comparison with an un-altered Paschen curve is presented in Fig. 14. The turning point disappears for the model with constant charge transfer reaction coefficient. The decrease of the fast-atom reflection coefficient with increasing $E/n$ (plotted in Fig. 6) also affects the onset of the runaway regime. The role of this parameter is rather similar to that of secondary electron yield at lower voltage, although ion-neutral ionization avalanches initiated by backscattered fast atoms actually propagate towards the cathode. The result of a calculation with a constant value of the fast-atom reflection coefficient, set at $R_a = 0.05$, is also shown in Fig. 14. The turning point in this example is at about $V_{br} = 600$ kV, much higher than 200 kV predicted by the un-altered model. With the reflection coefficient falling off at high $E/n$, the discharge gap-to-multiplication-length ratio needs to increase (while the multiplication length is also increasing). This results in the observed "C" shape of the Paschen curve, i. e., double-valued behavior, at very high $E/n = V/nd$.

## 4. Summary

We developed a reduced flux-balance analytical model to investigate ionization breakdown in helium for the applied voltage in the range of 10-1000 kV, with corresponding reduced density $nd \sim 1.5 \times 10^{20} m^{-2}$. In our electron/ion/fast neutral model, anisotropic scattering in gas-phase collisions and energy-dependent interactions at the electrode surfaces are carefully taken into account. Three regimes of the breakdown kinetics, labeled "electron regime", "ion regime", and "fast-neutral regime" are identified according to which species contributes the most to the gap-integrated ionization rate. In the fast-neutral regime of interest here, the Paschen curve for helium predicted by the model has been compared to that based on PIC/MCC simulations and to a set of experimental data, both presented in our preceding publication [11]. The Paschen curve predicted by this model is found to be in good agreement with the PIC/MCC result and with experimental data, also reported in [11]. Calculated profiles of particle fluxes in excellent agreement with those obtained in kinetic Monte Carlo simulations. Also, several underlying physical phenomena have been uncovered that are essential in electrical breakdown in extremely high electric field:

1. The significant role of fast neutral atoms, attributed to backscattering from the cathode and to impact ionization, is demonstrated by the analytical model. Fast-neutral backscattering from the cathode, rather similar in its function of ion-induced secondary electron emission at lower voltage, results in an ion-neutral avalanche, a self-organization mechanism which sustains the discharge current.

2. Stripping loss of fast neutrals, which is the primary loss channel under the condition of strongly anisotropic scattering in gas-phase collisions, has been proven essential to obtain a good agreement with experimental and PIC/MCC results. This is an indirect validation of the model which assumes velocity distributions strongly peaked in the direction of the electric field.

3. Lastly, the turning-point phenomenon, i. e., double-valued behavior of the Paschen currve, seen in the PIC/MCC model [11], is also predicted by the present analytical model. The nature of the turning point is that at extremely high values of reduced electric field *E/n* heavy species velocity distributions undergo a transition to runaway regime. This behavior is accounted for in the charge-transfer reaction rate of the reduced model. The turning point occurs primarily because the charge-transfer cross-section decreases with increasing projectile energy. The effect is amplified by the decrease in the fast-neutral flux reflection coefficient at high *E/n*.

**Acknowledgements**

The authors are indebted to Dr. Dmytro Sydorenko, the developer of the EDIPIC code. The work of L. Xu was supported by the Chinese Scholarship Council and the work of other authors by the U.S. Department of Energy, Office of Fusion Energy Sciences, and the Advanced Research Projects Agency-Energy (ARPA-E), under Award Number DE-AR0000298.

**Appendix A: Rate coefficients for helium**

In this appendix, we address energy-dependent values and corresponding integrated rates (reaction coefficients) for the cross-sections, secondary electron yields, and surface backscattering yields for ions and fast neutrals for discharge in helium in extremely high electric fields. The same data on cross-sections, electron yields, and reflection coefficients as adopted in Ref. [11] are used presently in order to make a valid comparison between the analytical and PIC/MCC kinetic models. The electron-impact ionization rate was discussed in the main text and will not be considered in this Appendix. In what follows, the values of reaction coefficients $\alpha_x / n$ are in units of $10^{-20} m^2$, cross sections $Q_x$ are in $10^{-20} m^2$, and particle energies $\varepsilon$ and effective ion temperature $T_i$ are in eV, unless noted otherwise. The coefficients given below depend on E/n through the effective ion temperature $T_i$ parameterizing velocity distribution, approximated with one-dimensional half-Maxwellian. The relation between $T_i$ and *E/n* is given by Eq. (7) in the main text.

**1. Reaction coefficients for energetic ions**

Approximate formula for charge-exchange cross section can be expressed as $Q_{ct}(\varepsilon) = (5.282 - 0.294 \ln \varepsilon)^2$, based on which the charge-exchange coefficient, obtained by flux-averaging over the ion distribution, can be in turn approximated as

$$\frac{\alpha_{ct}}{n}(T_i) = (5.282 - 0.294 \ln T_i)^2 + 0.171 + 0.339 \times (5.282 - 0.294 \ln T_i). \tag{1a}$$

For ion-impact ionization, we utilize the following analytic fit for the cross section data in the relevant energy range:

$$Q_{ii}(\varepsilon) = 0.252 + 1.099 \times 10^3 \sqrt{\varepsilon} / (\varepsilon + 4.650 \times 10^5). \tag{2a}$$

Therefore corresponding reaction coefficient is obtained by flux-averaging the ionization cross-section over the ion distribution. An analytical fit can be provided in the following form:

$$\frac{\alpha_{ii}}{n}(T_i) = \frac{10^{-4}}{T_i}\left[2.516\times 10^{-17}T_i + 1.099\times 10^{-13}\times\sqrt{\pi T_i} - 2353.172\times A'(a)\right] \quad (3a)$$

Where $A'(a) = \chi(a-100)\times(1-1/2a)/\sqrt{a\pi} + \chi(100-a)\times erf(\sqrt{a})\times\exp[a\times\chi(100-a)]$, $\chi$ is the step function, $erf$ is error function, and $a = 4.650\times 10^5 / T_i$. Note that $\frac{\alpha_{ii}}{n}$ in Eq. (3a) is in units of m$^2$.

## 2. Fast neutral atom reaction coefficients

### a. Primary fast atom-neutral ionization coefficient

For primary fast atoms produced in charge-exchange collisions, the ionization coefficient is expressed by Eq. (11) in the main text. To evaluate this expression, we first approximate the product of ionization cross section $Q_{ai}(\varepsilon)$ and charge exchange cross section $Q_{ct}(\varepsilon)$ as follows:

$$Q_{ct}(\varepsilon)Q_{ai}(\varepsilon) = 30.6685\times(\varepsilon/60842)/(\varepsilon/60842+1)^2, \quad (4a)$$

which is in unit of $10^{-40}\,m^4$. The following approximation can be given for the resulting value of $\frac{\alpha_{ai}}{n}(T_i)$ in Eq. (11):

$$\frac{\alpha_{ai}}{n}(T_i) = 30.6685\times b\times\left[(1+b)\times B'(b) - 1\right] \bigg/ \frac{\alpha_{ct}}{n} \quad (5a)$$

where $b = 60842/T_i$ and

$$B'[b(T_i)] = \exp[b\times\chi(1-b)]\times\chi(b)\times\chi(1-b)\times C'(b) + \chi(b-1)\times D'(b)/b,$$

$$C'[b(T_i)] = -0.577 + b\times\{1 + b\times\{-0.250 + b\times[0.055 + b\times(-0.010 + b\times 0.001)]\}\} - \ln(b),$$

$$D'[b(T_i)] = \frac{\{0.268 + b \times \{8.635 + b \times [18.059 + b \times (8.573 + b)]\}\}}{\{3.958 + b \times \{21.1 + b \times [25.633 + b \times (9.573 + b)]\}\}} .$$

**b. Ionization coefficient for backscattered fast neutrals.**

Eq. (13) in the main text gives an expression for the ionization coefficient due to fast neutrals backscattered from the cathode and travelling towards the anode. The denominator on the right-hand –side (first term) can be approximated as

$$\delta_2(T_i) = \{[3.7 - 0.315 \times \ln(T_i)]^2 + 0.196 + 0.364 \times [3.7 - 0.315 \times \ln(T_i)]\} \times T_i \qquad (6a)$$

and the numerator as

$$\delta_1(T_i) = 21550.225 \times \{-c \times \{[-1 + B'(c)c] + B'(c)\}\} + 1 - B'(c)c , \qquad (7a)$$

where $c(T_i) = 16641.1/T_i$ and the function $B'$ are the same as in Eq. (5a). Hence Eq. (13) yields

$$\frac{\alpha_{bi}}{n}(T_i) = \frac{\delta_1}{\delta_2} \times 1.5 . \qquad (8a)$$

**c. Secondary electron yields and fast-atom backscattering coefficient**

An analytical fit of the same form as developed in Ref. [14] for argon is used to approximate secondary electron yields of ions and of fast neutrals at the cathode. When averaged over the ion distribution $\exp(-\varepsilon/T_i)$, the electron emission yield by ions is approximated as

$$\gamma_i(T_i) = 0.3 + 1.55 \times (T_i/1000)^{1.9} / [1 + (T_i/600)^{1.54}] . \qquad (9a)$$

Likewise, by averaging over the resulting energy distribution of fast helium atoms produced in charge exchange collisions, the fast-atom induced electron emission yield is given by

$$\gamma_a(T_i) = 3 \times 10^{-5} \times (0.75 T_i)^{1.8} \times \exp(-622.5/T_i) / [1 + (T_i/400)^{1.56}] . \qquad (10a)$$

Regarding the backscattering coefficient $R_a$ of fast atoms at the cathode, it is convenient to convert Eq. (14) into the following form:

$$R_a(T_i) = \frac{\delta_2}{T_i} \bigg/ \frac{\alpha_{cx}}{n}. \tag{11a}$$

We note that the ion reflection (which includes neutralization) is neglected, owing to the ion flux collected at the cathode being much smaller than that of fast neutrals. Their ratio scales as $d/\lambda_{cx}$.

**Appendix B: Solution for the structure of the discharge.**

In this section, we present the solution to the linear equation comprising the present model of steady-state Townsend discharge (refer to model 4 in Table 1). It is given by the electron flux $\Gamma_e(x)$, the ion flux $\Gamma_i(x)$, the primary fast atom flux $\Gamma_a(x)$, and the reflected-fast-atom flux $\Gamma_b(x)$. According to Eqs. (3), (5), (10), and (12) with boundary conditions (8), (9) and (15), we find the particle fluxes (normalized to $\Gamma_t = \Gamma_e + \Gamma_i$) as follows:

$$\Gamma_e(x)/\Gamma_t = \left\{ \frac{[-NR_a/U_3 + (\beta + QR_a) \times U_2/U_1 - S]}{MR_a} \exp(\lambda_1 x) + \frac{1}{U_3}\exp(\lambda_2 x) - \frac{U_2}{U_1}\exp(\lambda_3 x) + 1 \right\} \bigg/ (1-P), \tag{1b}$$

$$\Gamma_i(x)/\Gamma_t = 1 - \Gamma_e(x)/\Gamma_t, \tag{2b}$$

$$\Gamma_a(x)/\Gamma_t = \left\{ -\frac{(-NR_a/U_3 + (\beta + QR_a) \times U_2/U_1 - S)}{R_a} \exp(\lambda_1 x) - \frac{N}{U_3}\exp(\lambda_2 x) + \frac{U_2}{U_1}Q\exp(\lambda_3 x) - S \right\} \bigg/ (1-P), \tag{3b}$$

$$\Gamma_b(x)/\Gamma_t = -\frac{U_2}{U_1}\beta \exp(\lambda_3 x) \bigg/ (1-P), \tag{4b}$$

where the functions M, N, Q, S, P and $\beta$ are those entering Eq. (16) in the main text and

$$U_1 = [\exp(\lambda_3 d) - \exp(\lambda_1 d) \times (\beta + QR_a)/MR_a] - \left[ \frac{\exp(\lambda_2 d) - \exp(\lambda_1 d)N/M}{(1+\gamma_i + N\gamma_a) - (1+\gamma_i + M\gamma_a)N/M} \right],$$
$$\times [(1+\gamma_i + Q\gamma_a) - (1+\gamma_i + M\gamma_a) \times (\beta + QR_a)/MR_a]$$

$$U_2 = [P - \exp(\lambda_1 d)S/MR_a] - \left[ \frac{\exp(\lambda_2 d) - \exp(\lambda_1 d)N/M}{(1+\gamma_i + N\gamma_a) - (1+\gamma_i + M\gamma_a)N/M} \right].$$
$$\times [(1+P\gamma_i + S\gamma_a) - (1+\gamma_i + M\gamma_a)S/MR_a]$$

$$U_3 = \frac{(1+\gamma_i + N\gamma_a) - (1+\gamma_i + M\gamma_a)N/M}{\{[(1+\gamma_i + Q\gamma_a) - (1+\gamma_i + M\gamma_a) \times (\beta + QR_a)/MR_a]U_2/U_1 - [(1+P\gamma_i + S\gamma_a) - (1+\gamma_i + M\gamma_a)S/MR_a]\}}$$

Alongside the full analytical model formulated above, the following three truncated models in Table 1 are introduced to aid in the discussion:

1. **Reduced model for electron regime (model 1):** $\alpha_{ii} = \alpha_{ai} = \gamma_a = R_a = 0$.

    These assumptions bring the original Townsend electron-multiplication model

    $$\frac{d\Gamma_{e,1}(x)}{dx} = \alpha_{ei}(E,n)\Gamma_{e,1}(x). \tag{5b}$$

    The boundary condition involves ion-induced electron emission from the cathode:

    $$\Gamma_{e,1}(0) = \gamma_i \Gamma_{i,1}(0) = \gamma_i \left[\Gamma_t - \Gamma_{e,1}(0)\right]. \tag{6b}$$

    Therefore, normalized electron and ion fluxes write as

    $$\frac{\Gamma_{e,1}(x)}{\Gamma_t} = \frac{\gamma_i}{\gamma_i + 1} \exp(\alpha_{ei} x), \tag{7b}$$

    $$\frac{\Gamma_{i,1}(x)}{\Gamma_t} = 1 - \frac{\Gamma_{e,1}(x)}{\Gamma_t}. \tag{8b}$$

2. **Second reduced model for ion regime (model 2):** $\alpha_{ai} = \gamma_a = R_a = 0$.

    Under this condition, the model simplifies as

    $$\frac{d\Gamma_{e,2}(x)}{dx} = -\frac{d\Gamma_{i,2}(x)}{dx} = \alpha_{ei}(E,n)\Gamma_{e,2}(x) + \alpha_{ii}(E/n)\Gamma_{i,2}(x), \tag{9b}$$

    with the same boundary condition Eq. (6b). The solution for electron and ion fluxes is

    $$\frac{\Gamma_{e,2}(x)}{\Gamma_t} = \left\{-\frac{1+\gamma_i \alpha_{ei}/\alpha_{ii}}{1+\gamma_i}\exp\left[(\alpha_{ei}-\alpha_{ii})x\right]+1\right\}\bigg/(1-\alpha_{ei}/\alpha_{ii}), \tag{10b}$$

$$\frac{\Gamma_{i,2}(x)}{\Gamma_t} = 1 - \frac{\Gamma_{e,2}(x)}{\Gamma_t} \quad . \tag{11b}$$

3. **Third reduced model for fast-neutral regime (model 3):** $R_a = 0$

   In the third model, Eqs. (3), (5) and (10) reduce to the following expressions:

$$-\frac{d\Gamma_{i,3}}{dx} = \frac{d\Gamma_{e,3}}{dx} = \alpha_{ei}\Gamma_{e,3} + \alpha_{ii}\Gamma_{i,3} + \alpha_{ai}\Gamma_{a,3}, \tag{12b}$$

$$\frac{d\Gamma_{a,3}}{dx} = -\alpha_{ct}\Gamma_{i,3}, \tag{13b}$$

with the boundary condition given by Eq. (8) in the main text:

$$\Gamma_{e,3}(0) = \gamma_i \Gamma_{i,3}(0) + \gamma_f \Gamma_{a,3}(0) \tag{14b}$$

$$\Gamma_{i,3}(d) = 0 \quad . \tag{15b}$$

Therefore we obtain the solution for the fluxes:

$$\frac{\Gamma_{e,3}(x)}{\Gamma_t} = \frac{-\exp[(\lambda_2 - \lambda_1)d] \times [1 + \gamma_a \alpha_{ei}/\alpha_{ai}]}{\{(1+\gamma_i)\{\exp[(\lambda_2 - \lambda_1)d] - 1\}\}} \exp(\lambda_1 x) + \frac{[1 + \gamma_a \alpha_{ei}/\alpha_{ai}]}{\{(1+\gamma_i)\{\exp[(\lambda_2 - \lambda_1)d] - 1\}\}} \exp(\lambda_2 x) + 1 \tag{16b}$$

$$\frac{\Gamma_{i,3}(x)}{\Gamma_t} = 1 - \frac{\Gamma_{e,3}(x)}{\Gamma_t} \quad . \tag{17b}$$